\documentclass[12pt,superscriptaddress,aps,prd,preprint,showpacs]{revtex4}
\usepackage{graphicx}
\usepackage{amsfonts}
\usepackage{slashed}
\usepackage[utf8x]{inputenc}
\usepackage{amsmath}
\usepackage{hyperref}

\def\bs{b\!\!\!/}

\begin{document}

\title{Carroll–Field–Jackiw term in a massless Rarita-Schwinger model}

\author{M. Gomes}
	\email{mgomes@if.usp.br}
\affiliation{Instituto de F\'\i sica, Universidade de S\~ao Paulo\\
Caixa Postal 66318, 05315-970, S\~ao Paulo, SP, Brazil}

\author{J. G. Lima}
\email{grimario.lima@fis.ufal.br}
\affiliation{Instituto de F\'\i sica, Universidade Federal de Alagoas,\\ 57072-900, Macei\'o, Alagoas, Brazil}

\author{T. Mariz}
\email{tmariz@fis.ufal.br}
\affiliation{Instituto de F\'\i sica, Universidade Federal de Alagoas,\\ 57072-900, Macei\'o, Alagoas, Brazil}

\author{J. R. Nascimento}
\email{jroberto@fisica.ufpb.br}
\affiliation{Departamento de F\'\i sica, Universidade Federal da Para\'\i ba,\\
 Caixa Postal 5008, 58051-970, Jo\~ao Pessoa, Para\'\i ba, Brazil}

\author{A. Yu. Petrov}
\email{petrov@fisica.ufpb.br}
\affiliation{Departamento de F\'\i sica, Universidade Federal da Para\'\i ba,\\
 Caixa Postal 5008, 58051-970, Jo\~ao Pessoa, Para\'\i ba, Brazil}

\begin{abstract}
We consider the massless Rarita-Schwinger (RS) LV QED. In this theory, we introduce the gauge fixing to obtain the propagator for the RS field, and calculate the Carroll-Field-Jackiw term, which turns out to be finite and ambiguous, and only in one calculation scheme, based on the nonlinear gauge framework, the gauge independence of the result is achieved.
\end{abstract}


\maketitle

\section{Introduction}

The formulation of the Lorentz-violating Standard Model extension (LV SME) in \cite{ColKost1,ColKost2} became a milestone that defines the framework for  studies of Lorentz-violating (LV) field theory models. In the mentioned papers, the minimal LV extensions of Abelian and non-Abelian gauge theories coupled to spinor and scalar fields were introduced, and the perturbative calculation of the Carroll-Field-Jackiw (CFJ) term was performed. Further extension of the LV SME has been carried out through the introduction of non-minimal operators, such as non-renomalizable couplings and higher-derivative kinetic terms \cite{KosLi1}. At the same time, one more direction for constructing a more generic LV field theory model consists in  the introduction of new fields. In this context, the most interesting candidate to be included in the extended model is the Rarita-Schwinger (RS), spin-3/2 field originally proposed in \cite{RS} and actively studied within the supergravity context (see f.e. \cite{Nieu}). In the Lorentz-invariant case, various issues related to the quantum dynamics of this field have been discussed in \cite{Pill,Pascal}, and simplest LV extensions of this theory together with its coupling to the gauge field, were constructed in \cite{prev}, where the examples of one-loop calculations for the case of the massive RS field, including the explicit obtention of the CFJ term, were performed. In \cite{prev1}, this theory was generalized for a non-Abelian gauge field.

In the massless case, the studies of quantum dynamics of the RS field become more complicated. Indeed, in this case the free RS theory turns out to possess a gauge symmetry, which requires the introduction of a gauge-fixing Lagrangian for the RS field, therefore its propagator must be strongly modified. Certainly, it will imply in very different results for quantum corrections in the massless RS field theory. In this case we will study this theory and calculate the lower LV quantum correction in the gauge sector, that is, the CFJ term. It should be emphasized that, since the propagators in the massless and massive cases have essentially distinct forms, the results for the CFJ term in the massless case can be different from those ones obtained in \cite{prev,prev1}, and in this paper we will demonstrate this difference explicitly.

The structure of the paper looks as follows. In section 2, we write down our Lagrangian and the corresponding Feynman rules. In the section 3 we calculate the one-loop contribution to the CFJ term for the linear gauge condition, and in the section 4, the same correction is obtained with use of the nonlinear gauge. Finally, in section 5, we discuss our results.

\section{Lagrangian and Feynman Rules}

The starting point is the massless Rarita-Schwinger Lagrangian of the spin-3/2 field coupled to the Maxwell field with the inclusion of a Lorentz-breaking term proportional to the constant axial vector $b^\mu$, given by (cf. \cite{prev1}):
\begin{equation}
{\cal L}_0=\bar{\psi}_{\mu}\frac{i}{2}\{\sigma^{\mu\nu},(i\slashed{\partial}-e\slashed{A}-\slashed{b}\gamma_5)\}\psi_{\nu}.
\end{equation}
This theory will be called the Rarita-Schwinger LV QED.

The above Lagrangian can be rewritten as
\begin{eqnarray}
{\cal L}_0&=&\bar{\psi}_{\mu}(i\slashed{D}g^{\mu\nu} 
-i(\gamma^{\mu}D^{\nu}+\gamma^{\nu}D^{\mu}) +i\gamma^{\mu}\slashed{D}\gamma^{\nu} +\slashed{b}\gamma_5 g^{\mu\nu} \nonumber\\
&&-(\gamma^{\mu}b^{\nu}+\gamma^{\nu}b^{\mu})\gamma_5 +\gamma^{\mu}\slashed{b}\gamma^{\nu}\gamma_5)\psi_{\nu},
\end{eqnarray}
where $D_\mu=\partial_\mu+ ie A_\mu$ is the covariant derivative. Observe that with $b=0$, the free part of the above expression is the form of the free spin-$3/2$ Lagrangian, corresponding to the choice of the constant $A$ defined in~\cite{Pill,prev} to be equal to 1, given by
\begin{equation}\label{LRS}
{\cal L}_{RS}=\bar{\psi}_{\mu}(i\slashed{\partial}g^{\mu\nu} 
-i(\gamma^{\mu}\partial^{\nu}+\gamma^{\nu}\partial^{\mu}) +i\gamma^{\mu}\slashed{\partial}\gamma^{\nu})\psi_{\nu}.
\end{equation}
It is also easy to observe that the above expression can be rewritten as
\begin{equation}
\label{LRS1}
{\cal L}_{RS}=-\epsilon^{\mu\nu\kappa\lambda}\bar{\psi}_{\mu}\gamma_5\gamma_\kappa\partial_\lambda\psi_{\nu}.
\end{equation}

Let us now consider the Feynman rules. The tensor operator of Lagrangian (\ref{LRS}) is not invertible  because of the gauge symmetry of the theory (see \cite{Das:1976ct} for details). From the form (\ref{LRS1}) of this Lagrangian, the invariance under the transformations $\delta\psi_{\nu}=\partial_{\nu}\chi$ folllows immediately. Therefore, it is necessary to add a gauge fixing term, e.g., given by
\begin{equation}
{\cal L}_{GF}=-\frac{1}{\alpha}\bar{\psi}_{\mu}\gamma^\mu i\slashed{\partial}\gamma^\nu\psi_{\nu}.\label{gaugefix}
\end{equation}
Thus, by considering that ${\cal L}_{RS}+{\cal L}_{GF}=\bar\psi_\mu (G^{-1})^{\mu\nu}\psi_\nu$ and $G_{\mu\alpha}(G^{-1})^{\alpha\nu}=i{g_\mu}^\nu$, for the spin-$3/2$ field propagator $G^{\mu\nu}$ in $D$ dimensions, we find
\begin{equation}
i G^{\mu\nu}(p)=\frac{i}{p^2}\left(\slashed{p}g^{\mu\nu}-\frac{2}{D-2}(\gamma^{\mu}p^{\nu}+\gamma^{\nu}p^{\mu})+\frac{1}{D-2}\gamma^\mu\slashed{p}\gamma^\nu\right) +\frac{i}{p^4}\left(\frac{4}{D-2}-\alpha\right)p^\mu\slashed{p}p^\nu,
\end{equation}
We note that the theory possesses two gauge symmetries, so that within the first of them, the standard one, the $A_{\mu}$ is transformed through the standard gauge transformations $A_{\mu}\to A_{\mu}+\partial_{\mu}\xi$, and the RS field plays the role of the matter transforming like $\psi_{\nu}\to e^{ie\xi}\psi_{\nu}$, and within the second one, the $\psi_{\mu}$ is transformed through its gradient gauge transformations $\delta\psi_{\nu}=\partial_{\nu}\chi$. At the same time, the gauge fixing term (\ref{gaugefix}) is evidently not invariant under the transformations $\psi_{\nu}\to e^{ie\xi}\psi_{\nu}$, i.e. besides the desired breaking of the gradient symmetry for $\psi_{\mu}$, necessary to define its propagator, it also breaks the standard gauge symmetry for this field, which is a less desired effect. It is clear that it could be very interesting to maintain the standard gauge symmetry for the RS field, but this will require a $A_{\mu}$-dependent gauge condition, i.e. a nonlinear gauge \cite{Adler1,Adler2}. We will consider such a possibility in the section IV. Now, as a first step  we consider a simplified situation, our gauge is linear, and there is no additional vertices generated by the gauge condition, but the consequence of this gauge choice will consist in gauge dependence of our CFJ term, as we will see further. We note that the propagator of the massive RS field \cite{prev}, in the zero mass limit, does not reduce to this expression in any gauge. It can be emphasized also that our theory is essentially four-dimensional, which excludes the scenario $D\to 2$ and the related singularity of the propagator.
For the RS-photon vertex, we write
\begin{equation}\label{ver}
-ie \left(g^{\mu\nu}\gamma^\lambda -(\gamma^{\mu}g^{\nu\lambda}+\gamma^{\nu}g^{\mu\lambda}) +\gamma^\mu\gamma^\lambda\gamma^\nu\right) = -ie \gamma^{\mu\lambda\nu}.
\end{equation}
\noindent
The coefficient for Lorentz and CPT violation $b_\mu$ leads to an insertion in the spin-3/2 field propagator, given by
\begin{equation}\label{ins}
-i (g^{\mu\nu}\slashed{b}\gamma_5 -(\gamma^{\mu}b^{\nu}+\gamma^{\nu}b^{\mu})\gamma_5 +\gamma^{\mu}\slashed{b}\gamma^{\nu}\gamma_5) = -i b_{\lambda}\gamma^{\mu\lambda\nu}\gamma_5,
\end{equation}
This insertion is the analogue of the standard $\bs\gamma_5$ insertion present within the calculation of the CFJ term in the standard LV QED, see f.e. \cite{JK}.

\section{Two-point functions: linear gauge}

\begin{figure}[htbp]
  \includegraphics[width=4cm]{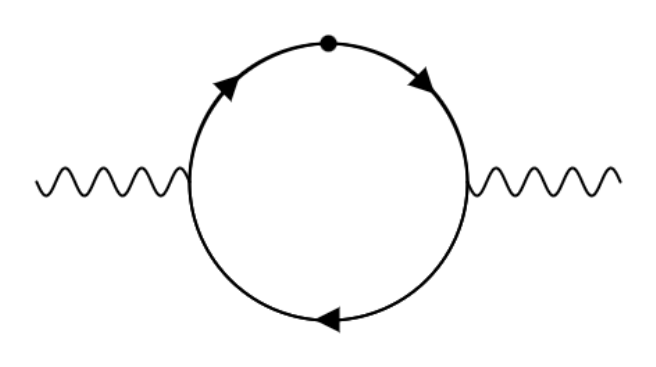}  
  \includegraphics[width=4cm]{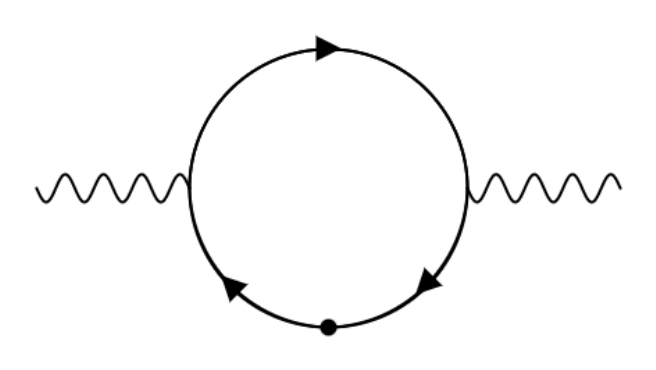}    
\caption{CFJ contributions to the two-point function of the vector field.}
\label{fig:1}
\end{figure}

Let us write down the contributions to the two functions of the gauge field, in which the graphs are depicted in Fig.~\ref{fig:1}, with the symbol $\bullet$ is for the insertion (\ref{ins}).

The corresponding action, with one insertion of the coefficient $b_\mu$ in the propagator $G^{\mu\nu}(p)$, is given by
\begin{equation}
S^{(2)}_{CFJ} = \frac{i}{2} \int \frac{d^4k}{(2\pi)^4} (\Pi_1^{\lambda\tau}+\Pi_2^{\lambda\tau})A_{\lambda}(-k)A_{\tau}(k),
\end{equation}
with
\begin{subequations}
\begin{eqnarray}
\label{twomin}
\Pi_1^{\lambda\tau} &=& e^2\,{\rm tr}\int\frac{d^4p}{(2\pi)^4}
\gamma^{\mu\lambda\nu} G_{\nu\alpha}(p) b_\kappa\gamma^{\alpha\kappa\beta}\gamma_5 G_{\beta\rho}(p) \gamma^{\rho\tau\sigma} G_{\sigma\mu}(p+k),\\
\label{Pi2}\Pi_2^{\lambda\tau} &=& e^2\,{\rm tr}\int\frac{d^4p}{(2\pi)^4}
\gamma^{\mu\lambda\nu} G_{\nu\rho}(p) \gamma^{\rho\tau\sigma} G_{\sigma\alpha}(p+k) b_\kappa\gamma^{\alpha\kappa\beta}\gamma_5 G_{\beta\mu}(p+k).
\end{eqnarray}
\end{subequations}
Note that, by considering the shift $p_\mu \to p_\mu-k_\mu$ in (\ref{Pi2}), we have
\begin{eqnarray}
\Pi_2^{\lambda\tau} &=& e^2\,{\rm tr}\int\frac{d^4p}{(2\pi)^4}
\gamma^{\rho\tau\sigma}G_{\sigma\alpha}(p) b_\kappa\gamma^{\alpha\kappa\beta}\gamma_5 G_{\beta\mu}(p)\gamma^{\mu\lambda\nu} G_{\nu\rho}(p-k) \nonumber\\
&=& \Pi_1^{\tau\lambda}(-k).
\end{eqnarray}
So, let us calculate only $\Pi_1^{\lambda\tau}$.

Thus, by considering $G^{\mu\nu}=G_1^{\mu\nu}+G_2^{\mu\nu}$, where
\begin{eqnarray}
G_1^{\mu\nu}(p)&=&\frac{1}{p^2}\left(\slashed{p}g^{\mu\nu}-\frac{2}{D-2}(\gamma^{\mu}p^{\nu}+\gamma^{\nu}p^{\mu})+\frac{1}{D-2}\gamma^\mu\slashed{p}\gamma^\nu\right), \\
G_2^{\mu\nu}(p)&=&\frac{1}{p^4}\left(\frac{4}{D-2}-\alpha\right)p^\mu\slashed{p}p^\nu,
\end{eqnarray}
we can rewrite (\ref{Pi2}) as $\Pi_1^{\lambda\tau}=\Pi_{1a}^{\lambda\tau}+\Pi_{1b}^{\lambda\tau}+\Pi_{1c}^{\lambda\tau}+\Pi_{1d}^{\lambda\tau}+\Pi_{1e}^{\lambda\tau}+\Pi_{1f}^{\lambda\tau}+\Pi_{1g}^{\lambda\tau}+\Pi_{1h}^{\lambda\tau}$, with
\begin{subequations}\label{Pi1t}
\begin{eqnarray}
\Pi_{1a}^{\lambda\tau} &=& e^2\,{\rm tr}\int\frac{d^4p}{(2\pi)^4}
\gamma^{\mu\lambda\nu} G_{1\nu\alpha}(p) b_\kappa\gamma^{\alpha\kappa\beta}\gamma_5 G_{1\beta\rho}(p) \gamma^{\rho\tau\sigma} G_{1\sigma\mu}(p+k), \\
\Pi_{1b}^{\lambda\tau} &=& e^2\,{\rm tr}\int\frac{d^4p}{(2\pi)^4}
\gamma^{\mu\lambda\nu} G_{1\nu\alpha}(p) b_\kappa\gamma^{\alpha\kappa\beta}\gamma_5 G_{1\beta\rho}(p) \gamma^{\rho\tau\sigma} G_{2\sigma\mu}(p+k), \\
\Pi_{1c}^{\lambda\tau} &=& e^2\,{\rm tr}\int\frac{d^4p}{(2\pi)^4}
\gamma^{\mu\lambda\nu} G_{1\nu\alpha}(p) b_\kappa\gamma^{\alpha\kappa\beta}\gamma_5 G_{2\beta\rho}(p) \gamma^{\rho\tau\sigma} G_{1\sigma\mu}(p+k), \\
\Pi_{1d}^{\lambda\tau} &=& e^2\,{\rm tr}\int\frac{d^4p}{(2\pi)^4}
\gamma^{\mu\lambda\nu} G_{1\nu\alpha}(p) b_\kappa\gamma^{\alpha\kappa\beta}\gamma_5 G_{2\beta\rho}(p) \gamma^{\rho\tau\sigma} G_{2\sigma\mu}(p+k), \\
\Pi_{1e}^{\lambda\tau} &=& e^2\,{\rm tr}\int\frac{d^4p}{(2\pi)^4}
\gamma^{\mu\lambda\nu} G_{2\nu\alpha}(p) b_\kappa\gamma^{\alpha\kappa\beta}\gamma_5 G_{1\beta\rho}(p) \gamma^{\rho\tau\sigma} G_{1\sigma\mu}(p+k), \\
\Pi_{1f}^{\lambda\tau} &=& e^2\,{\rm tr}\int\frac{d^4p}{(2\pi)^4}
\gamma^{\mu\lambda\nu} G_{2\nu\alpha}(p) b_\kappa\gamma^{\alpha\kappa\beta}\gamma_5 G_{1\beta\rho}(p) \gamma^{\rho\tau\sigma} G_{2\sigma\mu}(p+k), \\
\Pi_{1g}^{\lambda\tau} &=& e^2\,{\rm tr}\int\frac{d^4p}{(2\pi)^4}
\gamma^{\mu\lambda\nu} G_{2\nu\alpha}(p) b_\kappa\gamma^{\alpha\kappa\beta}\gamma_5 G_{2\beta\rho}(p) \gamma^{\rho\tau\sigma} G_{1\sigma\mu}(p+k), \\
\Pi_{1h}^{\lambda\tau} &=& e^2\,{\rm tr}\int\frac{d^4p}{(2\pi)^4}
\gamma^{\mu\lambda\nu} G_{2\nu\alpha}(p) b_\kappa\gamma^{\alpha\kappa\beta}\gamma_5 G_{2\beta\rho}(p) \gamma^{\rho\tau\sigma} G_{2\sigma\mu}(p+k).
\end{eqnarray}
\end{subequations}

\newpage

As the denominators are different, we must use different Feynman parametrizations (with the same shift, $p_\mu \to p_\mu+k_\mu(x-1)$), so that, after calculating the trace, we obtain
\begin{subequations}\label{Pi1t2}
\begin{eqnarray}
\Pi_{1a}^{\lambda\tau}&=&\int_0^1dx\, 2x\, \mu^{4-D} \int\frac{d^Dp}{(2\pi)^D}\frac{4 i e^2\epsilon^{\lambda\tau\mu\nu}b_\mu k_\nu}{(p^2-M^2)^3 (D-2)^2 D} ((D-2) ((D-5) D+2) D M^2 (x-1) \nonumber\\
&&+p^2 ((D-2) (D+2) ((D-5) D+2) x-2 (((D-9) D+22) D+28) D+304)),
\end{eqnarray}
\begin{eqnarray}
\Pi_{1b}^{\lambda\tau}&=& \int_0^1dx\, 6(x-1)x\, \mu^{4-D} \int\frac{d^Dp}{(2\pi)^D}\frac{4 i e^2\epsilon^{\lambda\tau\mu\nu}b_\mu k_\nu}{(p^2-M^2)^4(D-2)^2 D} (\alpha(D-2)-4) (k^2 p^2 x \nonumber\\
&&\times (D^2+2 (D-2) (D+4) x^2-4 ((D-3) D+6) x-4 D+16)+(D-2) \nonumber\\
&&\times D M^4 x +p^4 ((D-2) (D+4) x-2 (D-8) D-40)),
\end{eqnarray}
\begin{eqnarray}
\Pi_{1c}^{\lambda\tau}&=& -\int_0^1dx\, 3x^2\, \mu^{4-D} \int\frac{d^Dp}{(2\pi)^D}\frac{4 i e^2\epsilon ^{\lambda\tau\mu\nu}b_\mu k_\nu}{(p^2-M^2)^4(D-2) D} (\alpha(D-2)-4) (k^2 (x-1)^2 \nonumber\\
&&\times (D M^2 (x-1)+2 p^2 ((D+4) x-D+1))+p^4 (D (x-2)+4 x+2)),
\end{eqnarray}
\begin{eqnarray}
\Pi_{1d}^{\lambda\tau}&=& \int_0^1dx\, 12(x-1)x^2\, \mu^{4-D} \int\frac{d^Dp}{(2\pi)^D}\frac{4 i  e^2\epsilon^{\lambda\tau\mu\nu}b_\mu k_\nu}{(p^2-M^2)^5(D-2)^2 D (D+2)} p^2 (\alpha(D-2)-4)^2 \nonumber\\
&&\times (k^2 (p^2 (D^2 (x-1) (2 x+1)+3 D x+4 x (7-8 x))+(D+2) M^2 (x-1) \nonumber\\
&&\times ((D-4) x+D-1))+(D-4) (D+2) p^4),
\end{eqnarray}
\begin{eqnarray}
\Pi_{1e}^{\lambda\tau} &=& \int_0^1dx\, 3x^2\, \mu^{4-D} \int\frac{d^Dp}{(2\pi)^D}\frac{4 i e^2\epsilon^{\lambda\tau\mu\nu}b_\mu k_\nu}{(p^2-M^2)^4(D-2)^2 D} (\alpha(D-2)-4) (2 k^2 p^2 (x-1)^2 \nonumber\\
&&\times (-2 D^2+(3 D-14) (D+4) x+7 D+10)+(3 D-14) D k^2 M^2 (x-1)^3 \nonumber\\
&& +p^4 (D (3 D x-4 D-2 x+14)-56 x+20)),
\end{eqnarray}
\begin{eqnarray}
\Pi_{1f}^{\lambda\tau}&=& -\int_0^1dx\, 12(x-1)x^2\, \mu^{4-D} \int\frac{d^Dp}{(2\pi)^D}\frac{4 i e^2\epsilon^{\lambda\tau\mu\nu}b_\mu k_\nu}{(p^2-M^2)^5(D-2)^3 D (D+2) } \nonumber\\
&&\times p^2 (\alpha(D-2)-4)^2 (k^2 p^2 (-(D+4) (D^2+D-14) x+2 (D-4) (D-2) (D+4) x^2 \nonumber\\
&&+((3 D-22) D+48) D)+(D+2) (k^2 M^2 (x-1) (D ((D-6) x+D-13) \nonumber\\
&&+8 x+30) +(D-4) (D-2) p^4)),
\end{eqnarray}
\begin{eqnarray}
\Pi_{1g}^{\lambda\tau}&=& 0,
\end{eqnarray}
\begin{eqnarray}
\Pi_{1h}^{\lambda\tau}&=& 0,
\end{eqnarray}
\end{subequations}
where $M^2=k^2(x-1)x$.

Then, after we calculate the momentum integrals and sum the contributions, we have 
\begin{eqnarray}
\Pi_{1}^{\lambda\tau} = (N_0+\alpha N_1+\alpha^2 N_2) \epsilon^{\lambda\tau\mu\nu}b_\mu k_\nu,
\end{eqnarray}
with
\begin{subequations}
\begin{eqnarray}
N_0 &=& -\int_0^1dx\, \frac{2^{1-D} \pi ^{-\frac{D}{2}} x e^2\mu^{4-D} M^{D-4}}{(D-2)^3} \Gamma \left(2-\frac{D}{2}\right) (D (-16 (D-4) (D-2) (D-1) \nonumber\\
&&\times x^2+2 ((D (D+4)-75) D+234) D x+(D (D (17-3 D)+26) \nonumber\\
&& -300) D-472 x+312)+16 (9 x+28)),
\end{eqnarray}
\begin{eqnarray}
N_1 &=& \int_0^1dx\, \frac{2^{-D} \pi ^{-\frac{D}{2}} x e^2\mu^{4-D} M^{D-4}}{(D-2)^2} \Gamma \left(2-\frac{D}{2}\right) (D (-8 (D-4) (D-2) (D-1) x^2 \nonumber\\
&& +2 ((7 D-45) D+66) D x-(D-4) D (13 D-62)+88 x-320) \nonumber\\
&& +16 (74-9 x)),
\end{eqnarray}
\begin{eqnarray}
N_2 &=& \int_0^1dx\, \frac{2^{-D} \pi ^{-\frac{D}{2}} x e^2\mu^{4-D} M^{D-4}}{(D-2)} \Gamma \left(3-\frac{D}{2}\right) (D (D (-2 D+10 x+7)-38 x+20) \nonumber\\
&&+28 (x-3)).
\end{eqnarray}
\end{subequations}

Let us now expand the above expressions around $D=4$, so that we obtain
\begin{subequations}
\begin{eqnarray}
N_0 &=& \int_0^1dx\,  \left[-\frac{5x(3 x-2) e^2}{2 \pi ^2 \epsilon'} +\frac{x (8 (5-6 x) x-11)e^2}{4 \pi ^2}\right],
\end{eqnarray}
\begin{eqnarray}
N_1 &=& \int_0^1dx\,  \left[\frac{3x(3 x-2) e^2}{2 \pi ^2 \epsilon'} +\frac{x (3 (8 x-11) x+8)e^2}{4 \pi ^2}\right],
\end{eqnarray}
\begin{eqnarray}
N_2 &=& \int_0^1dx\,  \frac{x (9 x-5)e^2}{8 \pi ^2},
\end{eqnarray}
\end{subequations}
where $\frac 1{\epsilon'}=\frac 1\epsilon-\ln\frac M{\mu'}$, with $\epsilon=4-D$ and $\mu'^2=4\pi\mu^2e^{-\gamma}$.

Finally, by calculating the $x$ integrals, we have 
\begin{eqnarray}\label{Pi1T}
\Pi_1^{\lambda\tau}&=&-\frac{5e^2}{4\pi^2} \epsilon^{\lambda\tau\mu\nu}b_\mu k_\nu +\frac{\alpha(\alpha-2)e^2}{16\pi^2} \epsilon^{\lambda\tau\mu\nu}b_\mu k_\nu.
\end{eqnarray}
Therefore, for $\alpha=2$ or $\alpha=0$, we have
\begin{equation}
\Pi_1^{\lambda\tau} = -\frac{5e^2}{4\pi^2} \epsilon^{\lambda\tau\mu\nu}b_\mu k_\nu,
\end{equation}
and at $\alpha=\frac{1}{20}(-1\pm\sqrt{21})$, our CFJ term vanishes. We note that the non-triviality of our result consists in the fact that the CFJ term is found to be gauge dependent. This fact, in a certain sense, can be treated as an analogue of the famous ambiguity of the CFJ term in usual LV QED. 

\section{Two-point functions: nonlinear gauge}

 The above situation can change if we consider a gauge-fixing term, corresponding to a nonlinear gauge, that is,
\begin{equation}
\label{nonlingf}
{\cal L}_{GF}=-\frac{1}{\alpha}\bar{\psi}_{\mu}\gamma^\mu i\slashed{D}\gamma^\nu\psi_{\nu}
-\frac{1}{\beta}\bar{\psi}_{\mu}\gamma^\mu \slashed{b}\gamma^\nu\gamma_5\psi_{\nu},
\end{equation}
where new vertices and the LV insertion are introduced (for the Lorentz-invariant case such a scenario was treated in \cite{Adler1,Adler2}). We note that new gauge-fixing term does not break the "usual" gauge symmetry within which the RS field is transformed as a matter.  Thus, within the expressions (\ref{ver}) and (\ref{ins}), we have now
\begin{equation}
\label{newmatr}
\gamma_a^{\mu\lambda\nu}=g^{\mu\nu}\gamma^\lambda -g^{\mu\lambda}\gamma^\nu -g^{\nu\lambda}\gamma^\mu +\gamma^\mu\gamma^\lambda\gamma^\nu \left(1-\frac{1}{\alpha}\right)
\end{equation}
and
\begin{equation}
\label{newmatr1}
\gamma_b^{\mu\lambda\nu}=g^{\mu\nu}\gamma^\lambda -g^{\mu\lambda}\gamma^\nu -g^{\nu\lambda}\gamma^\mu +\gamma^\mu\gamma^\lambda\gamma^\nu \left(1-\frac{1}{\beta}\right),
\end{equation}
respectively. Here, within the new gauge-fixing term (\ref{nonlingf}), in order to keep track from its Lorentz-invariant and LV parts, we introduced two gauge parameters $\alpha$ and $\beta$, and only at the final step we consider the "simple" case $\alpha=\beta$.  We note also that the non-linearity of the gauge will imply in arising a new contribution to the CFJ term from the ghosts sector, besides the expected one, generated by the RS sector. Let us obtain both these results.

\subsection{Contribution from the RS sector}

For the RS sector, we can follow the same lines as in the previous section. So, after calculating the trace in (\ref{Pi1t}), now considering $\gamma_a^{\mu\lambda\nu}$ in the vertices and $\gamma_b^{\mu\lambda\nu}$ in the insertion, and evaluating the momentum integrals, using the same Feynman parametrizations as in (\ref{Pi1t2}), we obtain
\begin{eqnarray}
\Pi_{1}^{\lambda\tau} = (\alpha^{-2} N_{-2}+\alpha^{-1} N_{-1}+N_0+\alpha N_1+\alpha^2 N_2+\alpha^{3} N_{3}) \epsilon^{\lambda\tau\mu\nu}b_\mu k_\nu,
\end{eqnarray}
with
\begin{subequations}
\begin{eqnarray}
N_{-2} &=& \int_0^1dx\, \frac{2^{3-D} \pi^{-\frac{D}{2}} xe^2 \mu^{4-D} M^{D-4}}{3  \beta  (D-2)^3} \Gamma \left(2-\frac{D}{2}\right) (D (D^3 (1-2 x)^4+2 (2 (2 x-5) x \nonumber\\
&&+1) (D-2 D x)^2+4 D (2 x (5-2 (x-2) x)-5) x-8 x (2 x ((2 x-7) x+10) \nonumber\\
&&-3))+32 x),
\end{eqnarray}
\begin{eqnarray}
N_{-1} &=& -\int_0^1dx\, \frac{2^{1-D} \pi^{-\frac{D}{2}} xe^2 \mu^{4-D} M^{D-4}}{3  \beta  (D-2)^3} \Gamma \left(2-\frac{D}{2}\right) (3 D^5 (1-2 x)^4-6 D^4 (1-2 x)^2 \nonumber\\ 
&&\times(4 \beta  (x-2)+7)-4 D^3 (84 \beta -6 \beta  (8 (x-8) x+55) x+4 ((x (15 x-22)-60) x \nonumber\\
&&+63) x-67)+8 D^2 (84 \beta +6 \beta  (2 x (x+35)-49) x+4 x (56-83 x)-65) \nonumber\\
&&+32 D x (-39 \beta +2 x (3 x^2-3 \beta  (x+11)-4 x+28)+29)+64 (39 \beta -29) x),
\end{eqnarray}
\begin{eqnarray}
N_{0} &=& \int_0^1dx\, \frac{2^{-1-D} \pi^{-\frac{D}{2}} xe^2 \mu^{4-D} M^{D-4}}{3  \beta  (D-2)^3} \Gamma \left(2-\frac{D}{2}\right) (3 D^6 (1-2 x)^4+2 D^5 (6 \beta  (11-2 x \nonumber\\
&&\times(8 (x-3) x+19))-(1-2 x)^2 (2 (6 x-19) x+49))-4 D^4 (3 (97 \beta -61) \nonumber\\
&&+x (-912 \beta +2 x (516 \beta +2 x (-48 \beta +15 x+50)-473)+753))+8 D^3 (405 \beta \nonumber\\
&&+x (-1281 \beta +2 x (750 \beta +x (-36 \beta +30 x+35)-526)+951)-297)+16 D^2 \nonumber\\
&&\times(-111 \beta +x (519 \beta +2 x (-492 \beta +6 x^2+4 (7-6 \beta ) x+225)-601)+292) \nonumber\\
&&+32 D (-33 \beta +x (171 \beta -6 x (-38 \beta +x (-4 \beta +2 x+5)+10)+307)-160) \nonumber\\
&&-64 (21 \beta  (5 x+4)+82 x)),
\end{eqnarray}
\begin{eqnarray}
N_{1} &=& \int_0^1dx\, \frac{2^{-3-D} \pi^{-\frac{D}{2}} xe^2 \mu^{4-D} M^{D-4}}{3  \beta  (D-2)^2} \Gamma \left(2-\frac{D}{2}\right) (D^6 (-(1-2 x)^4)+4 D^5 (1-2 x)^2 \nonumber\\
&&\times(-12 \beta +2 x (3 \beta +x-8)+15)+8 D^4 (27 \beta +x (-207 \beta +2 x (2 (63 \beta -88) \nonumber\\
&&+x (-24 \beta +5 x+78))+233)-52)-16 D^3 (-57 \beta +2 x (-147 \beta +x (174 \beta \nonumber\\
&&+x (-9 \beta +5 x+43)-127)+118)-51)-16 D^2 (336 \beta +2 x (198 \beta +2 x \nonumber\\
&&\times(x^2-6 \beta  (x+18)+20 x-17)-205)+9)+64 D (6 \beta +2 x (45 \beta +x (x^2 \nonumber\\
&&-3 \beta  (x+8)+13 x-22)-132)+3)+256 (69 \beta +49 x-6)-2688 \beta  x),
\end{eqnarray}
\begin{eqnarray}
N_{2} &=& \int_0^1dx\, \frac{2^{-2-D} \pi^{-\frac{D}{2}} xe^2 \mu^{4-D} M^{D-4}}{3  \beta  (D-2)} \Gamma \left(2-\frac{D}{2}\right) (D (8 (D-5) (D-2) (D-1) (D+1) \nonumber\\
&&\times x^3 -4 (D-1) ((D (3 D-13)-20) D+80) x^2+2 (((3 D-17) (D+3) D+76) D \nonumber\\
&&+348) x -(D+7) ((D-8) (D+2) D+144))+6 \beta  (D (((2 D-15) D+8) D+164) \nonumber\\
&&-2 (D-4) (D-1) (5 D-14) x)-32 (63 \beta +23 x-66)),
\end{eqnarray}
\begin{eqnarray}
N_{3} &=& -\int_0^1dx\, \frac{2^{-2-D} \pi^{-\frac{D}{2}} xe^2 \mu^{4-D} M^{D-4}}{3  \beta} \Gamma \left(3-\frac{D}{2}\right) (D (3 D^2+4 (D-1) (D-3) x^2 \nonumber\\
&&-8 (D-3)^2 x-31 D+82)+8 (4 x-3)).
\end{eqnarray}
\end{subequations}

Finally, expanding the above expressions around $D=4$, we get
\begin{subequations}
\begin{eqnarray}
N_{-2} &=& \int_0^1dx\, \bigg[\frac{e^2}{\pi^2\beta\epsilon'} (8 (x-1) x (2 x-1) (15 (x-1) x+2) \nonumber\\
&& +\frac{e^2}{3\pi^2\beta} (x ((2 x (3 (66 x-175) x+470)-315) x+28))\bigg],
\end{eqnarray}
\begin{eqnarray}
N_{-1} &=& \int_0^1dx\, \bigg[-\frac{e^2}{\pi^2\beta\epsilon'}(2 x (8 \beta +x (-57 \beta +4 x (27 \beta +5 x (-3 \beta +9 x-20)+74) \nonumber\\
&& -81)+6)) +\frac{e^2}{3\pi^2\beta}(2 x (-12 \beta +x (69 \beta +x (-99 \beta +x (39 \beta -27 x+140) \nonumber\\
&& -148)+39)+1))\bigg],
\end{eqnarray}
\begin{eqnarray}
N_0 &=& \int_0^1dx\, \bigg[-\frac{e^2}{2\pi^2\beta\epsilon'} (x (-34 \beta +x (3 (71 \beta +40)-4 x (2 (51 \beta +62)+5 x (-12 \beta \nonumber\\
&& +18 x-37)))-8)) +\frac{e^2}{12\pi^2\beta}(x (-45 \beta +x (264 \beta +2 x (-318 \beta +x (204 \beta \nonumber\\
&& -486 x+569)+26)-411)+116))\bigg],
\end{eqnarray}
\begin{eqnarray}
N_1 &=& \int_0^1dx\, \bigg[-\frac{e^2}{2\pi^2\beta\epsilon'} (x (6 \beta +x (4 x (24 (\beta +1)+5 x (-3 \beta +3 x-7)) -9 (5 \beta +2)))) \nonumber\\
&&+\frac{e^2}{12\pi^2\beta} (x (42 \beta +2 x (-126 \beta +x (6 (40 \beta -19)+x (-141 \beta +171 x-149)) \nonumber\\
&&+120)-37))\bigg],
\end{eqnarray}
\begin{eqnarray}
N_2 &=& \int_0^1dx\, \bigg[-\frac{e^2}{\pi^2\beta\epsilon'} x^3 (5 x-4) +\frac{e^2}{24\pi^2\beta} (x (-15 \beta +x (27 \beta +2 x (56-13 x) \nonumber\\
&&-75)+2))\bigg],
\end{eqnarray}
\begin{eqnarray}
N_3 &=& -\int_0^1dx\, \frac{e^2}{4\pi^2\beta} x^3.
\end{eqnarray}
\end{subequations}

Then, by calculating the $x$ integrals, we obtain
\begin{equation}
N_{-2} = N_{-1} = 0
\end{equation}
and
\begin{equation}
N_0 = -\frac{5e^2}{4\pi^2},\ N_1 = \frac{e^2}{8\pi^2}+\frac{e^2}{4\pi^2\beta},\ N_2 = \frac{e^2}{16\pi^2}-\frac{e^2}{8\pi^2\beta},\ N_3 = -\frac{e^2}{16\pi^2\beta}, 
\end{equation}
so that we have
\begin{equation}\label{Pi1T2}
\Pi_1^{\lambda\tau} = \frac{e^2}{4\pi^2}\left(-5+\frac{\alpha}{2}+\frac{\alpha}{\beta}+\frac{\alpha^2}{4}-\frac{\alpha^2}{2\beta}-\frac{\alpha^3}{4\beta}\right) \epsilon^{\lambda\tau\mu\nu}b_\mu k_\nu,
\end{equation}
which perfectly corresponds to the CFJ term. We note the finiteness of this result. The case $\beta\to 0$ is excluded since it corresponds to the physically inconsistent scenario of a large LV term.
For $\beta\to\infty$, that is, if the gauge-fixing term is perfectly Lorentz-invariant and does not involve the LV vector, we obtain 
\begin{eqnarray}
\Pi_1^{\lambda\tau}&=&-\frac{5e^2}{4\pi^2} \epsilon^{\lambda\tau\mu\nu}b_\mu k_\nu +\frac{\alpha(\alpha+2)e^2}{16\pi^2} \epsilon^{\lambda\tau\mu\nu}b_\mu k_\nu,
\end{eqnarray}
which is only slightly different from (\ref{Pi1T}). Notably, for the "simple" case $\beta=\alpha$, the result (\ref{Pi1T2}) becomes surprisingly independent of the gauge parameter $\alpha$, leading to the gauge-invariant expression for the self-energy tensor
\begin{equation}
\Pi_1^{\lambda\tau} = -\frac{e^2}{\pi^2} \epsilon^{\lambda\tau\mu\nu}b_\mu k_\nu,
\end{equation}
therefore, in a certain sense, this result is preferable since it does not depend on any arbitrary parameters.

\subsection{Ghosts contribution}

As we noted above, non-linearity of our gauge corresponding to the gauge-fixing action (\ref{nonlingf}) naturally implies a nontrivial ghosts action.  Indeed, it is natural to present (\ref{nonlingf}), in the most interesting case $\alpha=\beta$,  as ${\cal L}_{GF}=-\frac{1}{\alpha}\bar{\psi}_{\mu}\gamma^\mu 
\chi$, where $\chi=(i\slashed{D}\gamma^\nu+ \slashed{b}\gamma^\nu\gamma_5)\psi_{\nu}$ is the gauge-fixing function.  Its variation under the standard gauge transformations of the RS field  whose form is $\delta\psi_{\nu}=\partial_{\nu}\lambda$, looks like $\delta\chi=(i\slashed{D}\gamma^\nu+ \slashed{b}\gamma^\nu\gamma_5)\partial_{\nu}\lambda$, which allows to write the Faddeev-Popov (FP) Lagrangian, following the standard prescription, in the form
\begin{equation}
{\cal L}_{FP}=-ic^{\prime}\delta\chi|_{\lambda\to c},
\end{equation}
where $c,c^{\prime}$ are the FP ghosts. Taking into account that $D_{\mu}=\partial_{\mu}+ieA_{\mu}$, we find
\begin{eqnarray}
{\cal L}_{FP}&=&-ic^{\prime}(i\slashed{D}\gamma^\nu+ \slashed{b}\gamma^\nu\gamma_5)\partial_{\nu}c \nonumber\\
&=&-c^{\prime}(-\gamma^{\mu}(\partial_{\mu}+ieA_{\mu})\gamma^\nu+ i\slashed{b}\gamma^\nu\gamma_5)\partial_{\nu}c \nonumber\\
&=&c^{\prime}(\Box+ie\gamma^{\mu}\gamma^{\nu}A_{\mu}\partial_{\nu}-i\slashed{b}\gamma^{\nu}\gamma_5\partial_{\nu})c.
\end{eqnarray}
Hence we have the ghosts propagator
\begin{equation}
<c^{\prime}(-p)c(p)>=\frac{i}{p^2-\slashed{b}\slashed{p}\gamma_5}=G_{gh}(p),
\end{equation}
and the vertex is $V=iec^{\prime}\gamma^{\mu}\gamma^{\nu}A_{\mu}\partial_{\nu}c$.

So, the simplest two-point diagram with a ghost loop looks like (the overall sign + is caused by $i^2$ from each of vertices, and $(-1)$ arising since ghosts are fermions):
\begin{eqnarray}
\Sigma_{gh}(k)=\frac{e^2}{2}{\rm tr}\int\frac{d^4p}{(2\pi)^4}\gamma^{\mu}\gamma^{\nu}A_{\mu}(-k)p_{\nu}G_{gh}(p)\gamma^{\alpha}\gamma^{\beta}A_{\alpha}(k)(p_{\beta}+k_{\beta})G_{gh}(p+k).
\end{eqnarray}
The explicit form of this expression is
\begin{eqnarray}
\Sigma_{gh}(k)&=&-\frac{e^2}{2}{\rm tr}\int\frac{d^4p}{(2\pi)^4}\gamma^{\mu}\gamma^{\nu}A_{\mu}(-k)p_{\nu}\frac{1}{p^2-\slashed{b}\slashed{p}\gamma_5}
\gamma^{\alpha}\gamma^{\beta}A_{\alpha}(k)(p_{\beta}+k_{\beta})\nonumber\\
&&\times\frac{1}{(p+k)^2-\slashed{b}(\slashed{p}+\slashed{k})\gamma_5}.
\end{eqnarray}
To get the CFJ contribution, we should keep in this expression only the first order in $b_{\mu}$. Taking into account only relevant terms, we arrive at
\begin{eqnarray}
\Sigma_{gh}(k)&=&-\frac{e^2}{2}{\rm tr}\int\frac{d^4p}{(2\pi)^4}\gamma^{\mu}\gamma^{\nu}A_{\mu}(-k)p_{\nu}\frac{\slashed{b}\slashed{p}\gamma_5}{p^4}
\gamma^{\alpha}\gamma^{\beta}A_{\alpha}(k)(p_{\beta}+k_{\beta})\frac{1}{(p+k)^2} \nonumber\\
&&-\frac{e^2}{2}{\rm tr}\int\frac{d^4p}{(2\pi)^4}\gamma^{\mu}\gamma^{\nu}A_{\mu}(-k)p_{\nu}\frac{1}{p^2}
\gamma^{\alpha}\gamma^{\beta}A_{\alpha}(k)(p_{\beta}+k_{\beta})\frac{\slashed{b}(\slashed{p}+\slashed{k})\gamma_5}{(p+k)^4}.
\end{eqnarray}
It remains to calculate traces and integrals. We immediately can see that 
\begin{eqnarray}
\Sigma_{gh}(k)=-\frac{e^2}{2}A_{\mu}(-k)A_{\alpha}(k)[T^{\mu\alpha}_1(k)+T^{\alpha\mu}_2(k)],
\end{eqnarray}
where $T^{\mu\alpha}_1(k)$ and $T^{\alpha\mu}_2(k)$, after a cyclic permutation of Dirac matrices, are given by the expressions
\begin{eqnarray}
T^{\mu\alpha}_1(k)&=&{\rm tr}\int\frac{d^4p}{(2\pi)^4}\gamma^{\mu}\gamma^{\nu}p_{\nu}\frac{\slashed{b}\slashed{p}\gamma_5}{p^4}
\gamma^{\alpha}\gamma^{\beta}(p_{\beta}+k_{\beta})\frac{1}{(p+k)^2},\nonumber\\
T^{\alpha\mu}_2(k)&=&{\rm tr}\int\frac{d^4p}{(2\pi)^4}\gamma^{\alpha}\gamma^{\beta}(p_{\beta}+k_{\beta})\frac{\slashed{b}(\slashed{p}+\slashed{k})\gamma_5}{(p+k)^4}\gamma^{\mu}\gamma^{\nu}p_{\nu}\frac{1}{p^2}.
\end{eqnarray}
In $T^{\alpha\mu}_2(k)$, we can make the change $p_\mu \to p_\mu-k_\mu$, so that we obtain
\begin{eqnarray}
T^{\alpha\mu}_2(k)&=&{\rm tr}\int\frac{d^4p}{(2\pi)^4}\gamma^{\alpha}\gamma^\beta p_\beta \frac{\slashed{b}\slashed{p}\gamma_5}{p^4}\gamma^{\mu}\gamma^{\nu}(p_{\nu}-k_{\nu})\frac{1}{(p-k)^2},
\end{eqnarray}
where, as a result, we observe that $T^{\mu\alpha}_1(k)=T^{\alpha\mu}_2(-k)$. This relation between two contributions is a standard situation for two contributions to the CFJ term, formed by LV insertions in two different propagators (see e.g.~\cite{ColKost2}). Therefore, let us calculate one of these contributions, say $T^{\mu\alpha}_1(k)$.
First, we use Dirac matrices commutators to write $\gamma^\nu p_\nu\slashed{b}\slashed{p}=2(p\cdot b)\slashed{p}-\slashed{b}p^2$. Thus, we have
\begin{eqnarray}
T^{\mu\alpha}_1(k)&=&{\rm tr}\int\frac{d^4p}{(2\pi)^4}\gamma^{\mu}[2(p\cdot b)\slashed{p}-\slashed{b}p^2]
\gamma^{\alpha}\gamma^{\beta}(p_{\beta}+k_{\beta})\gamma_5\frac{1}{p^4(p+k)^2}.
\end{eqnarray}
Now, we can calculate the trace of the product of Dirac matrices by the rule ${\rm tr}(\gamma^{\alpha}\gamma^{\beta}\gamma^{\gamma}\gamma^{\delta}\gamma_5)=4i\epsilon^{\alpha\beta\gamma\delta}$ and then obtain
\begin{eqnarray}\label{T1}
T^{\mu\alpha}_1(k)&=&4i\epsilon^{\mu\rho\alpha\sigma}\int\frac{d^4p}{(2\pi)^4}[2(p\cdot b)p_{\rho}(p_{\sigma}+k_{\sigma})-b_{\rho}(p_{\sigma}+k_{\sigma})p^2]\frac{1}{p^4(p+k)^2}.
\end{eqnarray}

After we employ the Feynman representation,
\begin{equation}
\frac{1}{p^4(p+k)^2}=2\int_0^1\frac{dx(1-x)}{[(p+kx)^2+k^2x(1-x)]^3},
\end{equation}
and define the new integration variable $l_\mu=p_\mu+k_\mu x$, we can rewrite (\ref{T1}) as
\begin{eqnarray}\label{T12}
T^{\mu\alpha}_1(k)&=&8i\epsilon^{\mu\rho\alpha\sigma}\int\frac{d^4l}{(2\pi)^4}
\int_0^1\frac{dx(1-x)}{[l^2+k^2x(1-x)]^3}[2(l\cdot b-k\cdot b x)(l_{\rho}-k_{\rho}x)(l_{\sigma}+k_{\sigma}(1-x)) \nonumber\\
&&-b_{\rho}(l_{\sigma}+k_{\sigma}(1-x))(l^2-2l\cdot k x+k^2x^2)],
\end{eqnarray}
where we must keep only the first order in the external $k_\mu$:
\begin{eqnarray}
\label{firstorder}
T^{\mu\alpha}_1(k)&=&8i\epsilon^{\mu\rho\alpha\sigma}\int\frac{d^4l}{(2\pi)^4}
\int_0^1\frac{dx(1-x)}{[l^2+k^2x(1-x)]^3}[-2l_{\rho}l_{\sigma}(k\cdot b)x-2k_{\rho}xl_{\sigma}(l\cdot b) \nonumber\\
&&+2l_{\rho}(l\cdot b)k_{\sigma}(1-x)-b_{\rho}(-2l_{\sigma}(l\cdot k)x+k_{\sigma}(1-x)l^2)].
\end{eqnarray}
Now, we employ the averaging over directions by the rule $l_{\rho}l_{\sigma}\to\frac{1}{4}g_{\rho\sigma}l^2$, which yields (taking into account that $\epsilon^{\mu\rho\alpha\sigma}g_{\rho\sigma}=0$)
\begin{eqnarray}
T^{\mu\alpha}_1(k)&=&8i\epsilon^{\mu\rho\alpha\sigma}\int\frac{d^4l}{(2\pi)^4}
\int_0^1\frac{dx(1-x) l^2}{[l^2+k^2x(1-x)]^3} \nonumber\\
&&\times[-\frac{1}{2}k_{\rho}b_{\sigma}x+\frac{1}{2}k_{\sigma}b_{\rho}(1-x)+b_{\rho}(\frac{1}{2}k_{\sigma}x-k_{\sigma}(1-x))].
\end{eqnarray}
Thus, after straightforward simplifying, we find
\begin{eqnarray}
T^{\mu\alpha}_1(k)&=&8i\epsilon^{\mu\rho\alpha\sigma}k_{\rho}b_{\sigma}\int\frac{d^4l}{(2\pi)^4}
\int_0^1dx \frac{l^2}{[l^2+k^2x(1-x)+\mu^2]^3}(1-x)(1-3x),
\end{eqnarray}
where we introduced the IR regulator $\mu^2$ in the denominator. Let us keep only the lower (linear) order in $k_\mu$ in the above expression, so that we have
\begin{eqnarray}
T^{\mu\alpha}_1(k)&=&4i\epsilon^{\mu\rho\alpha\sigma}k_{\rho}b_{\sigma}\int\frac{d^4l}{(2\pi)^4}
\int_0^1dx(1-x)(1-3x)\frac{l^2}{[l^2+\mu^2]^3},
\end{eqnarray}
with $\int_0^1dx(1-x)(1-3x)=0$. Therefore, we conclude that the first-order contribution in $k_{\mu}$ (which is only necessary to us) is zero, i.e., $T^{\mu\alpha}_1(k)=0$. Similarly, $T_2^{\alpha\mu}(k)$ vanishes as well, and thus the ghosts contribution is zero.

At the same time, if we average over directions by the rule $l_{\rho}l_{\sigma}\to\frac{1}{D}g_{\rho\sigma}l^2$ in (\ref{T12}), with also $d^4l/(2\pi)^4 \to \mu^{4-D}d^Dl/(2\pi)^D$, we arrive at the same zero result for $T^{\mu\alpha}_1(k)$ in all orders of $k_{\mu}$. 

We note that this calculation can be easily generalized for the case $\alpha\neq \beta$, the only modification to do is to replace $b_{\mu}\to\frac{\alpha}{\beta}b_{\mu}$ throughout this subsection, and the final result for the CFJ contribution from the ghost sector evidently again will be zero.

\section{Summary}

We considered the model in which the usual gauge field is coupled to the massless RS term, obtained the propagators in this model, and calculated the one-loop CFJ term. We found that it is finite as occurs not only in the usual LV QED (see f.e. \cite{JK}), but also in the LV theory of the massive RS field coupled to the gauge one considered in our previous papers \cite{prev}. We note that our result, for the linear gauge condition, first, depends on the gauge parameter, which is a very rare situation in the purely gauge sector, especially for the two-point function, second, does not reproduce any of the results found in \cite{prev}. This situation can be treated as a natural analogue of the scenario taking place in the usual spinor LV QED where the results for the CFJ term for massive and massless fermions are essentially different \cite{Perez-Victoria:2001csb}. Then, we performed the calculation for the nonlinear gauge condition, and found that within this scenario, in the case where the gauge-fixing term is completely described by only one parameter, the CFJ term is gauge independent, which apparently signalizes that this scheme is the most appropriate.  We note that, in some sense, this difference of results for the CFJ term can be treated as a certain analogue of the famous ambiguity of the CFJ term taking place in the usual LV QED, and caused by the fact that the gauge invariance of the CFJ term itself, together with the requirement for the procedure of the regularization of the one-loop contribution to be compatible with the gauge symmetry are not sufficient to remove this ambiguity, in a total analogy with the usual spinor QED \cite{Perez-Victoria:2001csb,Altschul:2004gs}, where the CFJ term is also superficially divergent. Moreover, this ambiguity can be naturally expected to be also related to some anomaly, potentially being able to display a scenario similar to \cite{Jackiw:1999qq}. Besides that, it should be noted that, within a possible embedding of our study in a curved space-time, natural within the supergravity context, the ambiguity of the CFJ term could break the Bianchi identities \cite{Altschul:2003ce,Altschul:2019eip}, hence, in a certain sense, even in the flat space-time the zero result for the CFJ term is preferable. At the same time, we note that within our calculations, we employed one well-defined regularization, that is, dimensional reduction. However, we note that the ambiguity of the CFJ term in our theory is rather natural since, first, our calculations involve the $\gamma_5$ matrix known to be ambiguously defined within the dimensional regularization context \cite{Altschul:2003ce}, just as occurs in the usual LV QED. It is natural to expect that in our theory, similarly to the usual spinor QED, one can relate an ambiguity of the CFJ term with that one of the chiral current, in analogy with \cite{Chung:1999gg}. We plan to study the functional integral formullation and the chiral current ambiguity for this theory in forthcoming papers.

{\bf One more} further continuation of this study could consist in generalizing our calculations to the non-Abelian case and studying of anomalies for the Rarita-Schwinger LV QED. Also, we can study various issues like coupling of massless LV RS theory to gravity, which effectively allows to study a LV supergravity model, or implementing the finite temperature. We plan to perform these studies in forthcoming works.

{\bf Acknowledgments.}  This work was partially supported by Conselho Nacional de Desenvolvimento Cient\'\i fico e Tecnol\'ogico (CNPq). The work of A. Yu.\ P. has been partially supported by the CNPq project No. 303777/2023-0.

\end{document}